\documentclass[10pt,english]{article}

\usepackage{amssymb}
\usepackage{amsmath}
\usepackage{amscd}
\usepackage{latexsym}  
\usepackage{epsfig}
\usepackage{bbm}
\usepackage{revsymb}

\newtheorem{defi}{Definition}
\newtheorem{lemma}[defi]{Lemma}
\newtheorem{thm}[defi]{Theorem}
\newtheorem{cor}[defi]{Corollary}
\newtheorem{rem}[defi]{Remark}
\newtheorem{prop}[defi]{Proposition}
\newtheorem{exempel}[defi]{Example}

\newcommand{\qed}{\hfill $\Box$}
\newcommand{\tr}{{\operatorname{Tr}}}

\newcommand{\conv}{{\operatorname{conv}\,}}
\newcommand{\Dist}{{\operatorname{Distr}}}
\newcommand{\bra}[1]{{\langle{#1}|}}
\newcommand{\ket}[1]{{|{#1}\rangle}}
\newcommand{\ketbra}[1]{{\ket{#1}\!\bra{#1}}}
\newcommand{\C}{{\mathbbm{C}}}
\newcommand{\E}{{\mathbbm{E}}}

\newcommand{\fset}[1]{{\mathcal{#1}}}

\newcommand{\1}{{\openone}}

\newenvironment{expl}[1][{}]{\begin{exempel} {#1}\normalfont}{\end{exempel}}

\newlength{\blank}
\newlength{\equalsign}
\newenvironment{beweis}[1][{\hspace{-\blank}}]{{\noindent\emph{Proof~{#1}.\ }}}{\hfill $\Box$\vskip 0.5\baselineskip}

\begin{document}

\title{Commitment Capacity of\protect\\ Discrete Memoryless Channels}
\author{Andreas Winter\thanks{Department of Computer Science, University of Bristol, Merchant
Venturers Building, Woodland Road, Bristol BS8 1UB, United Kingdom. Email: {\tt winter@cs.bris.ac.uk}}\qquad{}%
Anderson C.~A.~Nascimento\thanks{Imai Laboratory, Information and Systems, Institute of
Industrial Science, University of Tokyo, 4--6--1 Komaba, Meguro--ku, Tokyo 153--8505, Japan.
Email: {\tt anderson@imailab.iis.u-tokyo.ac.jp}, {\tt imai@iis.u-tokyo.ac.jp}}\qquad{}%
Hideki Imai${}^\dagger$
}

\date{31st March 2003}

\maketitle

\begin{abstract}
  In extension of the bit commitment task and following work initiated by Cr\'{e}peau
  and Kilian, we introduce and solve the problem of characterising the optimal rate at
  which a discrete memoryless channel can be used for bit commitment. It turns out that
  the answer is very intuitive: it is the maximum equivocation of the channel
  (after removing trivial redundancy),
  even when unlimited noiseless bidirectional side communication is allowed.
  By a well--known reduction, this result provides a lower bound on the
  channel's capacity for implementing coin tossing, which we conjecture to be
  an equality.
  \par
  The method of proving this relates the problem to Wyner's wire--tap channel
  in an amusing way. We also discuss extensions to quantum channels.
\end{abstract}

\section{Introduction}
\label{sec:intro}
Chess masters Alice and Bob are playing for the world chess championship and,
after playing for several hours, realize that they will have to stop the game
and resume it on the next morning. However, a problem arises: if Alice plays
her turn before stopping the game, Bob will have the entire night to think of
his next move, giving him an unfair advantage. If Alice does not play, she
will have the entire night to thing of her move. How can they get out of this
problem?
\par
If there is a trusted referee, Alice can write down her move and put
it into an envelope and give it to the referee, who will announce it to Bob in
the next morning. As the referee is trusted, Alice will be unable to change
her move after writing it down, also Bob will be unable to learn Alice's move
before the next morning. Can Alice and Bob solve this problem without the help
of a trusted referee?
\par
To solve these kind of problems without the use of an
active trusted party, Blum introduced commitment schemes in~\cite{Blum}. In a
commitment scheme, Alice commits to an information by sending some piece of
information to Bob during a commit phase. Later on, she can unveil the
information she committed to by sending some opening information to Bob during
an unveiling (also called reveal) phase. The protocol is said to be
concealing if the information sent by Alice during the commit phase does not
help Bob to learn a non--negligible amount of information on the value Alice is
committing to. It is said to be binding if Alice is unable to commit to a
certain information (which is usually a string of bits) and later on unveil a
different one.
\par
Without any kind of computational assumptions and assuming noiseless
communications, commitment schemes are impossible (see e.g.~\cite{DKS};
the generalisation to quantum protocols is due to Mayers~\cite{Mayers}).
Therefore, research has mostly focused on schemes were the
receiver is computationally bounded (computationally concealing schemes) or
schemes were the sender is computationally bounded (computationally binding
schemes). Examples of computationally binding but unconditionally concealing
schemes are~\cite{Blum},~\cite{BCC},~\cite{Halevi} and~\cite{HM}. Examples of
computationally concealing but unconditionally binding schemes are~\cite{Naor}
and~\cite{STACS}.
\par
It is now known that
noise is a powerful resource for the implementation of cryptographic primitives:
it allows for the construction of \emph{information theoretically secure} cryptographic
protocols --- a task typically impossible without the noise, and in practice done
by relaxing to \emph{computational security}, assuming conjectures from
complexity theory.
\par
In his famous paper~\cite{Wyner:wiretap}, Wyner was the first to exploit noise in
order to establish a secure channel in the presence of an eavesdropper. These
results were extended in studies of secret key distillation by Maurer~\cite{Maurer},
Ahlswede and Csisz\'{a}r~\cite{AC1} and followers. The noise in these studies is
assumed to affect the eavesdropper: thus, to work in practice, it has to be
guaranteed or certified somehow. This might be due to some --- trusted --- third party
who controls the channel (and thus prevents the cryptographic parties from cheating), or
due to physical limitations, as in quantum key distribution~\cite{BB84,general:PA}.
Recently, Cr\'{e}peau and Kilian~\cite{crepeau:kilian} showed how information
theoretically secure bit commitment can be implemented using a binary symmetric channel,
their results being improved in~\cite{crepeau} and~\cite{DKS}.
\par
The object of the present study is to optimise the use of the noisy channel,
much as in Shannon's theory of channel capacities: while the previous studies have
concentrated on the \emph{possibility} of bit commitment using noisy channels,
here we look at committing to one out of a larger message set, e.g.~a bit string.
We are able, for a general discrete memoryless channel, to characterise the
commitment capacity by a simple (single--letter) formula (theorem~\ref{thm:main}),
stated in section~\ref{sec:formalities}, and proved in two parts in
sections~\ref{sec:lowerbound} and~\ref{sec:upperbound}. A few specific examples
are discussed in section~\ref{sec:examples} to illustrate the main result.
In section~\ref{sec:quantum} results on an extension to quantum channels are related,
and we close with a discussion (section~\ref{sec:discussion}). An appendix
collects some facts abut typical sequences used in the main proof.

\section{Definitions and main result}
\label{sec:formalities}
In the commitment of a message there are two parties, called \emph{Alice} and
\emph{Bob}, the first one given the message $a$ from a certain set ${\cal A}$. The whole procedure
consists of two stages: first the \emph{commit phase}, in which Alice (based on $a$) and Bob
exchange messages, according to a protocol. This will leave Bob with a record (usually called
\emph{view}), to be used in the second stage, the \emph{reveal phase}. This consists of
Alice disclosing $a$ and other relevant information to Bob. Bob performs a test
on all his recorded data which accepts if Alice followed the rules and disclosed the
correct information in the second stage, and rejects if a violation of the rules is
discovered.
\par
To be useful, such a scheme has to fulfill two requirements: it must be ``concealing''
as well as ``sound'' and ``binding'': the first property means that after the commit phase
Bob has no or almost no information about $a$ (i.e., even though Alice has ``committed''
herself to something by the communications to Bob, this commitment remains secret),
and this has to hold even if Bob does not follow the protocol, while Alice does.
Soundness means that if both parties behave according to the protocol, Bob's test
will accept (with high probability) after the reveal phase.
The protocol to be binding means that Bob's test is such that
whatever Alice did in the commit phase (with Bob following the rules) there is only at
most one $a$ she can ``reveal'' which passes Bob's test.
\par\medskip
In our present consideration there is an unlimited bidirectional noiseless channel
available between Alice and Bob, and in addition a discrete memoryless noisy channel
$W:\fset{X}\longrightarrow\fset{Z}$ from Alice to Bob, which may be used $n$ times:
on input $x^n=x_1\ldots x_n$, the output distribution on ${\cal Z}^n$ is
$W^n_{x^n}=W_{x_1}\otimes\cdots\otimes W_{x_n}$.
\begin{defi}
  \label{defi:non-redundant}
  The channel $W$ is called \emph{non--redundant}, if none of its output distributions
  is a convex combination of its other output distributions:
  $$\forall y\forall P\text{ s.t. }P(y)=0\quad W_y\neq \sum_x P(x) W_x.$$
  In geometric terms this means that all distributions $W_x$ are distinct extremal points
  of the polytope ${\cal W}=\conv\{W_x:x\in{\cal X}\}$, the convex hull of
  the output distributions within the probability simplex over ${\cal Z}$.
  Clearly, we can make $W$ into a non--redundant channel $\widetilde{W}$ by removing
  all input symbols $x$ whose output distribution $W_x$ is not extremal. The old channel
  can be simulated by the new one, because by feeding it distributions over input symbols
  one can generate the output distributions of the removed symbols.
  \par
  The channel $W$ is called \emph{trivial}, if after making it non--redundant
  its output distributions have mutually disjoint support. This means that from the
  output one can infer the input with certainty.
\end{defi}
\par
With this we can pass to a formal definition of a protocol: this, consisting of
the named two stages, involves creation on Alice's side of either messages intended
for the noiseless channel, or inputs to the noisy channel, based on previous
messages receive from Bob via the noiseless channel, which themselves are based on
data received before, etc. Both agents may employ probabilistic choices, which we
model by Alice and Bob each using a random variable, $M$ and $N$, respectively.
This allows them to use \emph{deterministic} functions in the protocol.
Note that this makes all messages sent and received into well-defined random
variables, \emph{dependent on $a$}.
\par\medskip\noindent
\emph{Commit Phase:} The protocol goes for $r$ rounds of Alice--to--Bob and
Bob--to--Alice noiseless communications $U_j$ and $V_j$.
After round $r_i$ ($r_1\leq\ldots\leq r_n\leq r$) Alice will also send a symbol
$X_i$ down the noisy channel $W$, which Bob receives as $Z_i$. Setting
$r_0=0$ and $r_{n+1}=r$:
\\
Round $r_i+k$ ($1\leq k\leq r_{i+1}-r_i$): Alice sends
$U^{r_i+k}=f_{r_i+k}\bigl(a,M,V^{r_i+k-1}\bigr)$ noiselessly.
Bob answers $V_{r_i+k}=g_{r_i+k}\bigl(Z^i,N,U^{r_i+k}\bigr)$,
also noiselessly. After round $r_i$ and before round $r_i+1$
($1\leq i\leq n$), Alice sends $X_i=F_i\bigl(a,M,V^{r_i}\bigr)$,
which Bob receives as $Z_i=W(X_i)$.
\par\medskip\noindent
\emph{Reveal Phase:} A similar procedure as the Commit Phase, but without
the noisy channel uses, including Alice's sending $a$ to Bob.
At the end of the exchange Bob performs a test as to whether to accept
Alice's behaviour or not. It is easily seen that this procedure can be
simulated by Alice simply telling Bob $a$ and $M$, after which
Bob performs his test $\beta\bigl(Z^n,N,U^r;a,M\bigr)\in\{{\rm ACC},{\rm REJ}\}$.
I.e., requiring Alice to reveal $M$ and $a$ makes cheating for her only
more difficult.
\par\medskip
We shall, for technical reasons, impose the condition that the range of
the variable $U^r$ is bounded:
\begin{equation}
  \label{eq:U:rate}
  \bigl| U^r \bigr| \leq \exp(Bn),
\end{equation}
with a constant $B$. Note that $\exp$ and $\log$ in this paper are always to
basis $2$, unless otherwise stated.
\par
Now, the mathematical form of the conditions for concealing as well as for
soundness and binding is this: we call the above protocol
\emph{$\epsilon$--concealing} if for any two messages $a,a'\in{\cal A}$
and any behaviour of Bob during the commit phase,
\begin{equation}
  \label{eq:concealing}
  \frac{1}{2}\bigl\| \Dist_a(Z^n N U^r) - \Dist_{a'}(Z^n N U^r) \bigr\|_1
                                                                  \leq \epsilon,\tag{A}
\end{equation}
where $\Dist_a(Z^nNU^r)$ is the distribution of the random variables $Z^nNU^r$
after completion of the commit phase which Alice entered with the message $a$
and the randomness $M$,
and with the $\ell_1$--norm $\|\cdot\|_1$; the above expression is identical to
the total variational distance of the distributions. This is certainly the strongest
requirement one could wish for: it says that no statistical test of Bob
immediately after the commit phase can distinguish between
$a$ and $a'$ with probability larger than $\epsilon$. Note that $V^r$ is a function of
$Z^n N U^r$, and hence could be left out in eq.~(A).
Assuming any probability distribution on the messages, $a$ is the value of a random
variable $A$, and it is jointly distributed with all other variables of the protocol.
Then, whatever Bob's strategy,
\begin{equation}
  \label{eq:weak:concealing}
  I\bigl(A\wedge Z^n N U^r\bigr)\leq
           \epsilon'=H(2\epsilon,1-2\epsilon)+2n\epsilon(\log B+\log|{\cal Z}|),\tag{A'}
\end{equation}
where
$$I(X\wedge Y) = H(X)+H(Y)-H(XY)$$
is the (Shannon) mutual information between $X$ and $Y$, and
$$H(X) = -\sum_x \Pr\{X=x\}\log\Pr\{X=x\}$$
is the (Shannon) entropy of $X$~\cite{Shannon:Comm}.
\par
We call the protocol \emph{$\delta$--sound and  --binding}
(\emph{$\delta$--binding} for short), if for
Alice and Bob following the protocol, for all $a\in{\cal A}$,
\begin{equation}
  \label{eq:sound}
  \Pr\bigl\{ \beta\bigl(Z^nNU^r;a M\bigr)={\rm ACC} \bigr\} \geq 1-\delta,\tag{B1}
\end{equation}
and, whatever Alice does during the commit phase, governed by a random variable $S$
with values $\sigma$ (which determines the distribution of $Z^nNU^r$), for
all $A=a(S,V^r)$, $A'=a'(S,V^r)$, $\widetilde{M}=\mu(S,V^r)$
and $\widetilde{M}'=\mu'(S,V^r)$ such that $A\neq A'$ with probability $1$,
\begin{equation}
  \label{eq:binding}
  \Pr\Bigl\{ \beta\bigl( Z^nNU^r;A\widetilde{M} \bigr)={\rm ACC}\ \&\
             \beta\bigl( Z^nNU^r;A'\widetilde{M}'\bigr)={\rm ACC} \Bigr\} \leq \delta.\tag{B2}
\end{equation}
\par
Note that by convexity the cheating attempt of Alice is w.l.o.g.~\emph{deterministic},
which is to say that $S$ takes on only one value $\sigma$ with non-zero probability,
hence $\Pr\{S=\sigma\}=1$.
\par
We call $\frac{1}{n}\log|{\cal A}|$ the \emph{(commitment) rate} of the protocol.
A rate $R$ is said to be \emph{achievable} if there exist commitment protocols
for every $n$ with rates converging to $R$, which are $\epsilon$--concealing
and $\delta$--binding with $\epsilon,\delta\rightarrow 0$ as $n\rightarrow\infty$.
The \emph{commitment capacity $C_{\rm com}(W)$} of $W$ is the supremum of all
achievable rates.
\par
The main result of this paper is the following theorem:
\begin{thm}
  \label{thm:main}
  The commitment capacity of the discrete channel $W$ (assumed to be non--redundant) is
  $$C_{\rm com}(W) = \max\bigl\{ H(X|Z) : X,Z\text{ RVs, }\Dist(Z|X)=W \bigr\},$$
  i.e., the maximal \emph{equivocation} of the channel over all possible input distributions.
\end{thm}
\begin{cor}
  \label{cor:BC}
  Every non--trivial discrete memoryless channel can be used to perform
  bit commitment. \qed
\end{cor}
By invoking the well--known reduction of coin tossing to bit commitment~\cite{Blum}
we obtain:
\begin{cor}
  \label{cor:coin-tossing}
  The channel $W$ can be used for secure two--party coin tossing at rate
  at least $C_{\rm com}(W)$. I.e., for the naturally defined coin tossing
  capacity $C_{\rm c.t.}(W)$, one has $C_{\rm c.t.}(W) \geq C_{\rm com}(W)$.  \qed
\end{cor}
This theorem will be proved in the following two sections
(propositions~\ref{prop:lower} and~\ref{prop:upper}): first we construct a protocol
achieving the equivocation bound, showing that exponential decrease of $\epsilon$ and
$\delta$ is possible, and without using the noiseless side channels at all during the commit
phase. Then we show the optimality of the bound.
\par\medskip
To justify our allowing small errors both in the concealing and the binding property of
a protocol, we close this section by showing that demanding too much trivialises the problem:
\begin{thm}
  \label{thm:zero-error}
  There is no bit--commitment via $W$ which is $\epsilon$--concealing and $0$--binding
  with $\epsilon<1$. I.e., not even two distinct messages can be committed: $|{\cal A}|=1$.
\end{thm}
\begin{beweis}
  If the protocol is $0$--sound, this means that
  for \emph{every} value $\mu$ attained by $M$ with positive probability,
  Bob will accept the reveal phase if Alice behaved according to the protocol.
  On the other hand, that the protocol is $0$--binding means that for
  $a\neq a'$ and arbitrary $\mu'$, Bob will never accept if Alice behaves
  according to the protocol in the commit phase, with values $a\mu$ but tries
  to ``reveal'' $a'\mu'$. This opens the possibility of a decoding method for
  $a$ based on $Z^nNU^r$: Bob simply tries out all possible $a\mu$ with his test
  $\beta$ --- the single $a$ which is accepted must be the one used by Alice.
  Hence the scheme cannot be $\epsilon$--concealing with $\epsilon<1$.
\end{beweis}
\begin{rem}
  \label{rem:zero-concealing}
  By contrast, is is easy to construct schemes both  $0$--concealing and
  $\delta$-binding with $\delta<1$, for an appropriately defined channel:
  \par
  Consider the channel $F$ with input and output alphabets
  ${\cal X}={\cal Z}=\{0,1,2,3\}$ and signals defined by
  \begin{equation*}
    F_x(z) =\begin{cases}
              \frac{1}{2} & \text{if }\ z-x\equiv 0\text{ or }1\mod 4, \\
              0           & \text{otherwise.}
            \end{cases}
  \end{equation*}
  Alice commits to the bit $b$ by picking $c\in\{0,1\}$ at random and sending
  $x=b+2c$ (for which Bob receives a random $z$ such that $z-x\equiv 0\text{ or }1\mod 4$).
  To reveal, she tells him $x$ (which decodes to a unique $b$)
  and he accepts iff $z-x\equiv 0\text{ or }1\mod 4$.
  \par
  Cleary, this scheme is $0$--concealing because for both $b=0$ and $b=1$
  Bob sees the uniform distribution on ${\cal Z}$. 
  Equally obviously, it is $0$--sound, but it is also $\frac{1}{2}$--binding:
  for if Alice wants to ``reveal'' $b'\neq b$ (with corresponding $x'\neq x$),
  she has only a probability of $\frac{1}{2}$ to pick $x'$ with
  $z-x'\equiv 0\text{ or }1\mod 4$.
\end{rem}
\par
This simple scheme is in fact the template for the coding scheme of
proposition~\ref{prop:lower}: put differently, on sufficiently large scale
(block length) every channel looks like $F$.

\section{A scheme meeting the equivocation bound}
\label{sec:lowerbound}
Here we describe and prove security bounds of a scheme which is very simple compared
to the generality we allowed in section~\ref{sec:upperbound}: in the commit phase
it consists only of a single block use of the noisy channel $W^n$, with no public
discussion at all, where the input $X^n$ is a random one--to--many function of $a$
(in particular, $a$ is a function of the $x^n$ chosen). In the reveal phase Alice simply
announces $X^n$ to Bob.
\begin{prop}
  \label{prop:code}
  Given $\sigma,\tau>0$, and a distribution $P$ of $X\in{\cal X}$,
  with the output $Z=W(X)$, $Q=\Dist(Z)$. Then there exists a collection of codewords
  $$\bigl( \xi_{a\mu}\in{\cal X}^n : a=1,\ldots,K,\ \mu=1,\ldots,L \bigr)$$
  with the following properties:
  \begin{enumerate}
    \item For all $(a,\mu)\neq(a',\mu')$, $d_{\rm H}(\xi_{a\mu},\xi_{a'\mu'}) \geq 2\sigma n$.
    \item For every $a$:
      $$\frac{1}{2}\left\| \frac{1}{L}\sum_{\mu=1}^L W^n_{\xi_{a\mu}}-Q^{\otimes n} \right\|_1
                                                          \leq 25|{\cal X}||{\cal Z}|\exp(-n\tau).$$
    \item There are constants $G,G'$ and a continuous function $G''$ vanishing at $0$
          such that
          \begin{align*}
            K &\geq \frac{1}{2n}(3+\log|{\cal X}|+\log|{\cal Z}|)^{-1}
                       \exp\bigl( n H(X|Z)-n\sqrt{2\tau}G'-n G''(\sigma) \bigr),                    \\
            L &\leq n(3+\log|{\cal X}|+\log|{\cal Z}|)\exp\bigl( n I(X\wedge Z)+n\sqrt{2\tau}G \bigr).
          \end{align*}
  \end{enumerate}
\end{prop}
\begin{beweis}
  To get the idea, imagine a wiretap channel~\cite{Wyner:wiretap}
  with $W$ as the stochastic matrix of the eavesdropper
  and a symmetric channel $S_\sigma:{\cal X}\longrightarrow{\cal Y}={\cal X}$
  for the legal user:
  \begin{equation*}
    S_\sigma(y|x) = \begin{cases}
                      1-\sigma                   & \text{ if }x=y,   \\
                      \frac{1}{|{\cal X}|}\sigma & \text{ if }x\neq y.
                    \end{cases}
  \end{equation*}
  The random coding strategy for such a channel, according to Wyner's solution~\cite{Wyner:wiretap}
  (but see also~\cite{csiszar:koerner} and~\cite{Cai:Yeung})
  will produce a code with the properties $2$ and $3$. Because the code for the legal user
  must fight the noise of the symmetric channel $S_\sigma$, we can expect its codewords to
  be of large mutual Hamming distance, i.e., we should get property $1$.
  \par
  In detail: pick the $\xi_{a\mu}$ i.i.d.~according to the distribution
  $\widetilde{P}^n$, which is $0$ outside ${\cal T}^n_{P,\sqrt{2\tau}}$
  (the typical sequences, see appendix A) and $P^{\otimes n}$ within, suitably normalised.
  Also introduce the subnormalised measures $\widehat{W}^n_{x^n}$: this is
  identical to $W^n_{x^n}$ within ${\cal T}^n_{W,\sqrt{2\tau}}(x^n)$ and $0$ outside.
  We will show that with high probability we can select codewords
  with properties 2 and 3, and only a small proportion of which violate
  property 1; then by an expurgation argument will we obtain the desired code.
  \par
  By eqs.~(\ref{eq:typ:probability}) and~(\ref{eq:c-typ:probability})
  in the appendix we have
  \begin{equation}
    \label{eq:approx-1}
    \frac{1}{2}\left\| \E\widehat{W}^n_{\xi_{a\mu}}-Q^{\otimes n} \right\|
                                                      \leq 3|{\cal X}||{\cal Z}|\exp(-n\tau),
  \end{equation}
  with the expectation referring to the distribution $\widetilde{P}^n$ of the
  $\xi_{a\mu}$.
  Observe that the support of all $\widehat{W}^n_{\xi_{a\mu}}$ is contained
  in ${\cal T}^n_{Q,2|{\cal X}|\sqrt{\tau}}$, using eq.~(\ref{eq:c-typ:in:typ})
  of the appendix.
  Now, ${\cal S}$ is defined as the set of those $z^n$ for which
  $$\E\widehat{W}^n_{\xi_{a\mu}}(z^n) \geq T:=\exp(-n\tau)
                                              \exp\bigl( -nH(Q)-n\sqrt{2\tau}|{\cal X}|F \bigr),$$
  with $F=\sum_{z:Q(z)\neq 0} -\log Q(z)$, and define
  $\widetilde{W}^n_{x^n}(z^n)=\widehat{W}^n_{x^n}(z^n)$ if $z^n\in{\cal S}$
  and $0$ otherwise. With the cardinality estimate eq.~(\ref{eq:typ:upper})
  of the appendix and eq.~(\ref{eq:approx-1}) we obtain
  \begin{equation}
    \label{eq:approx-2}
    \frac{1}{2}\left\| \E\widetilde{W}^n_{\xi_{a\mu}}-Q^{\otimes n} \right\|
                                                      \leq 4|{\cal X}||{\cal Z}|\exp(-n\tau).
  \end{equation}
  The Chernoff bound allows us now to efficiently sample
  the expectation $\widetilde{Q}^n:=\E\widetilde{W}^n_{\xi_{a\mu}}$:
  observe that all the values of $\widetilde{W}^n_{\xi_{a\mu}}$ are upper bounded by
  $$t:=\exp\bigl( -nH(W|P)+n\sqrt{2\tau}|{\cal X}|\log|{\cal Z}|+n\sqrt{2\tau} E \bigr),$$
  using eq.~(\ref{eq:c-typ:value}).
  Thus, rescaling the variables, by lemma~\ref{lemma:chernoff} and with the
  union bound, we get
  \begin{equation}\begin{split}
    \label{eq:approx-3}
    \Pr &\left\{ \forall a\forall z^n\in{\cal S}\
                   \frac{1}{L'}\sum_{\mu=1}^{L'} \widetilde{W}^n_{\xi_{a\mu}}(z^n)
                             \in \bigl[(1\pm\exp(-n\tau))\widetilde{Q}^n(z^n)]\right\}    \\
        &\phantom{=====================}
         \leq 2K|{\cal S}|\exp\left( -L\frac{T}{t}\frac{\exp(-2n\tau)}{2\ln 2}\right),
  \end{split}\end{equation}
  which is smaller than $1/2$ if
  $$L' > 2+n(\log|{\cal X}|+\log|{\cal Z}|)
               \exp\bigl( n I(X\wedge Z)+n\sqrt{2\tau}G \bigr),$$
  with $G=3+|{\cal X}|F+|{\cal X}|\log|{\cal Z}|+E$.
  Note that in this case, there exist values for the $\xi_{a\mu}$ such that the
  averages $\frac{1}{L'}\sum_\mu W^n_{\xi_{a\mu}}$ are close to $Q^{\otimes n}$.
  \par
  Now we have to enforce property 1: in a random batch of $\xi_{a\mu}$ we call
  $a\mu$ \emph{bad} if $\xi_{a\mu}$ has Hamming distance less than $2\sigma n$ from another
  $\xi_{a'\mu'}$. The probability that $a\mu$ is bad is easily bounded:
  \begin{equation*}\begin{split}
    \Pr\{a\mu\text{ bad}\} &\leq 2|{\cal X}|\exp(-n\tau)
                                +P^{\otimes n}\left(
                                         \bigcup_{a'\mu'\neq a\mu} B_{2n\sigma}(\xi_{a'\mu'}) \right)        \\
                           &\leq 2|{\cal X}|\exp(-n\tau)
                                +\max \left\{ P^{\otimes n}({\cal A}) :
                                               |{\cal A}| \leq K'L' {n \choose 2n\sigma} |{\cal X}|^{2n\sigma}
                                                                                                    \right\} \\
                           &\leq 5|{\cal X}|\exp(-n\tau),
  \end{split}\end{equation*}
  by eq.~(\ref{eq:shadow:lower}) in the appendix, because we choose
  \begin{equation*}\begin{split}
    K' &\leq \frac{1}{n}(3+\log|{\cal X}|+\log|{\cal Z}|)^{-1}              \\
       &\phantom{===}
             \exp\bigl( n H(X|Z)-n\sqrt{2\tau}G-2n\sqrt{2\tau}D
                        -n H(2\sigma,1-2\sigma)-2n\sigma\log|{\cal X}| \bigr),
  \end{split}\end{equation*}
  hence
  \begin{equation*}
    K'L' {n \choose 2n\sigma} |{\cal X}|^{2n\sigma}
                                 <\exp\bigl( nH(P)-2n\sqrt{2\tau}D \bigr).
  \end{equation*}
  Thus, with probability at least $1/2$, only a fraction of $10|{\cal X}|\exp(-n\tau)$
  of the $a\mu$ are bad. Putting this together with eq.~(\ref{eq:approx-3}), we obtain
  a selection of $\xi_{a\mu}$ such that
  \begin{equation}
    \label{eq:approx-4}
    \forall a\quad \frac{1}{2}\left\| \frac{1}{L'}\sum_\mu W^n_{\xi_{a\mu}}
                                                               -Q^{\otimes n} \right\|_1
                      \leq 5|{\cal X}||{\cal Z}|\exp(-n\tau)
  \end{equation}
  and only a fraction of $10|{\cal X}|\exp(-n\tau)$ of the $a\mu$ are bad.
  \par
  This means that for at least half of the $a$, w.l.o.g.~$a=1,\ldots,K=K'/2$,
  only a fraction $20|{\cal X}|\exp(-n\tau)$ of the $\mu$ form bad pairs $a\mu$,
  w.l.o.g.~for $\mu=L+1,\ldots,L'$, with $L=\bigl(1-20|{\cal X}|\exp(-n\tau)\bigr)L'$.
  Throwing out the remaining $a$ and the bad $\mu$, we are left with a
  code as desired.
\end{beweis}
\par
Observe that a receiver of $Z^n$ can efficiently check claims about the input
$\xi_{a\mu}$ because of property $1$, that distinct codewords have ``large''
Hamming distance. The non--redundancy of $W$ shuns one--sided errors in this checking,
as we shall see.
The test $\beta$ is straightforward: it accepts iff $Z^N\in{\cal T}^n_{W,\sqrt{2\tau}}(\xi_{a\mu})$,
the set of \emph{conditional typical sequences}, see appendix A. This ensures
soundness; for the bindingness we refer to the following proof.
\par
We are now in a position to describe a protocol, having chosen codewords
according to proposition~\ref{prop:code}:
\par\medskip\noindent
\begin{quote}
  \emph{Commit phase:} To commit to a message $a$, Alice picks $\mu\in\{1,\ldots,L\}$ uniformly at
    random and sends $\xi_{a\mu}$ through the channel. Bob obtains a channel output $z^n$.
  \par\medskip\noindent
  \emph{Reveal phase:} Alice announces $a$ and $\mu$. Bob performs the test $\beta$: he accepts
    if $z^n\in {\cal B}_{a\mu}:={\cal T}^n_{W,\sqrt{2\tau}}(\xi_{a\mu})$ and rejects otherwise.
\end{quote}
\par\medskip
\begin{prop}
  \label{prop:lower}
  Assume that for all $x\in{\cal X}$ and distributions $P$ with $P(x)=0$,
  $$\left\| W_x - \sum_y P(y)W_y \right\|_1 \geq \eta.$$
  Let $\tau=\frac{\sigma^4\eta^2}{8|{\cal X}|^4|{\cal Z}|^2}$:
  then the above protocol implements an $\epsilon$--concealing
  and $\delta$--binding commitment with rate
  $$\frac{1}{n}\log K \geq H(X|Z)-\sqrt{2\tau}G'
                            -H(2\sigma,1-2\sigma)-2\sigma\log|{\cal X}|
                            -\frac{\log n}{n}-O\left(\frac{1}{n}\right)$$
  and exponentially bounded security parameters:
  \begin{align*}
    \epsilon &= 50|{\cal X}||{\cal Z}|\exp(-n\tau),         \\
    \delta   &= 2|{\cal X}||{\cal Z}|\exp\bigl(-2n\tau^2\bigr).
  \end{align*}
\end{prop}
\begin{beweis}
  That the protocol is $\epsilon$--concealing is obvious from property 2 of the code
  in proposition~\ref{prop:code}: Bob's distribution of $Z^n$ is always $\epsilon/2$-close
  to $Q^{\otimes n}$, whatever $a$ is.
  \par
  To show $\delta$--bindingness observe first that if Alice is honest, sending $\xi_{a\mu}$
  in the commit phase and later revealing $a\mu$, the test $\beta$ will accept with
  high probability:
  \begin{equation*}\begin{split}
    \Pr\bigl\{ Z^n\in{\cal B}_{a\mu} \bigr\}
                       &=    W^n_{\xi_{a\mu}}\bigl( {\cal T}^n_{W,\sqrt{2\tau}}(\xi_{a\mu}) \bigr) \\
                       &\geq 1-2|{\cal X}||{\cal Z}|\exp(-n\tau)
                        \geq 1-\delta,
  \end{split}\end{equation*}
  by eq.~(\ref{eq:c-typ:probability}) in the appendix.
  \par
  On the other hand, if Alice cheats, we may --- in accordance with our definition ---
  assume her using a deterministic strategy: i.e., she ``commits'' sending some $x^n$
  and later attempts to ``reveal'' either $a\mu$ or $a'\mu'$, with $a\neq a'$.
  Because of property 1 of the code in proposition~\ref{prop:code}, at least one of the
  codewords $\xi_{a\mu}$, $\xi_{a'\mu'}$ is at Hamming distance at least $\sigma n$
  from $x^n$: w.l.o.g., the former of the two. But then the test $\beta$
  accepts ``revelation'' of $a\mu$ with small probability:
  \begin{equation*}
    \Pr\bigl\{ Z^n\in{\cal B}_{a\mu} \bigr\}
                       =    W^n_{x^n}\bigl( {\cal T}^n_{W,\sqrt{2\tau}}(\xi_{a\mu}) \bigr)
                       \leq 2\exp(-2n\tau^2) \leq \delta,
  \end{equation*}
  by lemma~\ref{lemma:distinct-types} in the appendix.
\end{beweis}

\section{Upper bounding the achievable rate}
\label{sec:upperbound}
We assume that $W$ is non--redundant. We shall prove the following assertion,
assuming a uniformly distributed variable $A\in{\cal A}$ of messages.
\begin{prop}
  \label{prop:upper}
  Consider an $\epsilon$--concealing and $\delta$--binding commitment protocol
  with $n$ uses of $W$. Then
  \begin{equation}\begin{split}
    \label{eq:bc-rate:upper}
    \log|{\cal A}|\leq &n\max\{H(X|Z):\Dist(Z|X)=W\}                              \\
                       &\phantom{==}+n\bigl( \epsilon(\log B+\log|{\cal Z}|)
                                            +5\sqrt[3]{\delta}\log|{\cal X}| \bigr)
                                    +2.
  \end{split}\end{equation}
\end{prop}
The key, as it turns out, of its proof, is the insight that in the
above protocol, should it be concealing and binding, $x^n$ together
with Bob's view of the commit phase (essentially)
determine $a$. In the more general formulation we permitted in
section~\ref{sec:formalities}, we prove :
\begin{equation}
  \label{eq:C:functionof:M}
  H(A|Z^n N U^r;X^n)\leq \delta'= H\left(5\sqrt[3]{\delta},1-5\sqrt[3]{\delta}\right)
                                 +5\sqrt[3]{\delta}\log|{\cal A}|.\tag{B'}
\end{equation}
Intuitively, this means that with the items Alice entered into the
commit phase of the protocol and those which are accessible to Bob,
not too many values of $A$ should be consistent ---
otherwise Alice had a way to cheat.
\par
\begin{beweis}[of eq.~(\ref{eq:C:functionof:M})]
  For each $a\mu$ the commit protocol (both players being honest) creates a distribution
  $\Delta_{a\mu}$ over \emph{conversations} $(x^n v^r;z^n u^r)$. We leave out Bob's random
  variable $N$ here, noting that he can create its correct conditional distribution
  from $z^n u^r;v^r$, which is his view of the conversation.
  The only other place where he needs it is to perform the test $\beta$.
  We shall in the following assume that it includes this creation of $N$,
  which makes $\beta$ into a probabilistic test, depending on
  $(a\mu v^r;z^n u^r)$.
  \par
  The pair $a\mu$ has a probability $\alpha_{a\mu}$ that its conversation with subsequent
  revelation of $a\mu$ is accepted. By soundness, we have
  $$\sum_\mu \Pr\{ M=\mu \}\alpha_{a\mu} \geq 1-\delta,$$
  for every $a$. Hence, by Markov inequality, there exists (for every $a$)
  a set of $\mu$ of total probability $\geq 1-\sqrt[3]{\delta^2}$ for which
  $\alpha_{a\mu}\geq 1-\sqrt[3]{\delta}$. We call such $\mu$ \emph{good for} $a$.
  \par
  From this we get a set ${\cal C}_{a\mu}$ of ``partial'' conversations $(x^n v^r;u^r)$,
  with probability $\Delta_{a\mu}\bigl({\cal C}_{a\mu}\bigr)\geq 1-\sqrt[3]{\delta}$,
  which are accepted with probability at least $1-\sqrt[3]{\delta}$.
  (In the test also $Z^n$ enters, which is distributed according to $W^n_{x^n}$.)
  \par
  Let us now define the set
  $${\cal C}_a := \bigcup_{\mu\text{ good for }a} {\cal C}_{a\mu},$$
  which is a set of ``partial conversations'' which are accepted with probability
  at least $1-\sqrt[3]{\delta}$ and
  $$\Delta_a\bigl( {\cal C}_a \bigr) \geq 1-2\sqrt[3]{\delta},$$
  with the distribution
  $$\Delta_a := \sum_\mu \Pr\{ M=\mu \}\Delta_{a\mu}$$
  over ``partial conversations'': it is the distribution created by the commit phase
  give the message $a$.
  \par
  We claim that
  \begin{equation}
    \label{eq:almost-done}
    \Delta_a\left( X^n V^r;U^r \in \bigcup_{a'\neq a} {\cal C}_{a'} \right)
                                                              \leq 3\sqrt[3]{\delta}.
  \end{equation}
  Indeed, if this were not the case, Alice had the following cheating strategy:
  in the commit phase she follows the protocol for input message $a$. In the
  reveal phase she looks at the ``partial conversation''
  $x^n v^r;u^r$ and tries to ``reveal'' some $a'\mu'$ for which
  the partial conversation is in ${\cal C}_{a'\mu'}$
  (if these do not exist, $a'\mu'$ is arbitrary).
  This defines random variables $A'$ and $\widetilde{M}'$ for which
  it is easily checked that
  $$\Pr\bigl\{ \beta(Z^n N U^r;a M)={\rm ACC}\ \&\
               \beta(Z^n N U^r;A'\widetilde{M}')={\rm ACC} \bigr\} > \delta,$$
  contradicting the $\delta$--bindingness condition.
  \par
  Using eq.~(\ref{eq:almost-done}) we can build a decoder for $A$ from $X^n V^r;U^r$:
  choose $\widehat{A}=a$ such that $X^n V^r;U^r\in{\cal C}_a$ --- if there exists none or
  more than one, let $\widehat{A}$ be arbitrary. Clearly,
  $$\Pr\{A\neq\widehat{A}\}\leq 5\sqrt[3]{\delta},$$
  and invoking Fano's inequality we are done.
\end{beweis}
\par
Armed with this, we can now proceed to the
\par\noindent
\begin{beweis}[of proposition~\ref{prop:upper}]
  We can successively estimate,
  \begin{equation*}\begin{split}
    H(X^n|Z^n) &\geq H(X^n|Z^n N U^r)                     \\
               &=    H(AX^n|Z^n N U^r)-H(A|Z^n N U^r;X^n) \\
               &\geq H(A|Z^n N U^r)   -H(A|Z^n N U^r;X^n) \\
               &\geq H(A|Z^n N U^r)   -\delta'            \\
               &=    H(A)-I(A\wedge Z^n N U^r)-\delta'    \\
               &\geq H(A)-\epsilon'   -\delta',
  \end{split}\end{equation*}
  using eq.~(B') in the fourth, eq.~(A') in the sixth line.
  On the other hand, subadditivity and the conditioning inequality imply
  $$H(X^n|Z^n)\leq \sum_{k=1}^n H(X_k|Z_k),$$
  yielding the claim, because $H(A)=\log|{\cal A}|$.
  \par
  The application to the proof of the converse of theorem~\ref{thm:main}
  is by observing that $\epsilon',\delta'=o(n)$.
\end{beweis}
\par\medskip
Note that for the proof of the proposition
we considered only a very weak attempt of Alice to cheat: she behaves
according to the protocol during the commit phase, and only at the reveal
stage she tries to be inconsistent. Similarly, our concealingness
condition considered only passive attempts to cheat by Bob, i.e., he
follows exactly the protocol, and tries to extract information about
$A$ only by looking at his view of the exchange.
\par
Thus, even in the model of \emph{passive cheating}, which is less restrictive
than our definition in section~\ref{sec:formalities}, we obtain the
upper bound of proposition~\ref{prop:upper}

\section{Examples}
\label{sec:examples}
In this section we discuss some particular channels, which we present
as stochastic matrices with the rows containing the output distributions.
\par\medskip\noindent
{\bf 1. Binary symmetric channel $B_p$:} Let $0\leq p\leq 1$. Define
\par
$B_p:=\quad$
\begin{tabular}{c||c|c}
  $\cdot$ & $0$   & $1$   \\
  \hline\hline
  $0$     & $1-p$ & $p$   \\
  \hline
  $1$     & $p$   & $1-p$
\end{tabular}
\par
The transmission capacity if this channel is easily computed from Shannon's
formula~\cite{Shannon:Comm}: $C(B_p)=1-H(p,1-p)$, which is non--zero
iff $p\neq 1/2$. The optimal input distribution is the uniform distribution
$(1/2,1/2)$ on $\{0,1\}$.
Note that this channel is trivial if $p\in\{0,1/2,1\}$, hence
$C_{\rm com}(B_p)=0$ for these values of $p$.
We may thus, w.l.o.g., assume that $0<p<1/2$,
for which $B_p$ is non--redundant. It is not hard to compute the optimal input
distribution as the uniform distribution, establishing $C_{\rm com}(B_p)=H(p,1-p)$.
\par
The result is in accordance with our intuition: the noisier the channel is, the worse
it is for transmission, but the better for commitment.
\par\medskip\noindent
{\bf 2. A trivial channel:} Consider the channel
\par
$T:=\quad$
\begin{tabular}{c||c|c}
  $\cdot$ & $0$   & $1$   \\
  \hline\hline
  $a$     & $1/2$ & $1/2$ \\
  \hline
  $b$     & $1$   & $0$   \\
  \hline
  $c$     & $0$   & $1$
\end{tabular}
\par
Clearly, $T$ is trivial, hence $C_{\rm com}(T)=0$. Still it is an interesting example in
the light of our proof of proposition~\ref{prop:lower}: for assume a wiretap channel
for which $T$ is the stochastic matrix of the eavesdropper, while the legal user obtains
a noiseless copy of the input. Then clearly the wiretap capacity of this system is
$1$, with optimal input distribution $(1/2,1/4,1/4)$.
\par\medskip\noindent
{\bf 3. Transmission and commitment need not be opposites:}
We show here an example of a channel where the optimising input distributions for transmission
and for commitment are very different:
\par
$V:=\quad$
\begin{tabular}{c||c|c}
  $\cdot$ & $0$   & $1$   \\
  \hline\hline
  $0$     & $1/2$ & $1/2$ \\
  \hline
  $1$     & $1$   & $0$
\end{tabular}
\par
It can be easily checked that the maximum of the mutual information, i.e.~the transmission
capacity, is attained for the input distribution
$$P(0)=\frac{2}{5}=0.4,\quad P(1)=\frac{3}{5}=0.6,$$
from which we obtain $C(V)\approx 0.3219$.
On the other hand, the equivocation is maximised for the input distribution
$$P'(0)=1-\sqrt{\frac{1}{5}}\approx 0.5528,\quad P'(1)=\sqrt{\frac{1}{5}}\approx 0.4472,$$
from which we get that $C_{\rm com}(V)\approx 0.6942$.
The maximising distributions are so different that the sum $C(V)+C_{\rm com}(V) > 1$,
i.e. it exceeds the maximum input and output entropies of the channel.

\section{Quantum channels}
\label{sec:quantum}
The construction of section~\ref{sec:lowerbound} can be carried over to
a class of quantum channels, namely so--called \emph{cq--channels}
(classical--quantum channels):
$$W:{\cal X} \longrightarrow {\cal S}({\cal H}),$$
a map from an input alphabet (here assumed to be finite) into the set of
states on a Hilbert space ${\cal H}$, also assumed to be finite--dimensional
in the present discussion.
(For an overview of quantum information theory see~\cite{bennett:shor}
and the textbook~\cite{nielsen:chuang}.)
\emph{Non--redundancy} means the same here, only that the convex structure is now the
convex compact set of states, instead of the probability simplex. We assume this property
of $W$ silently in the following.
\begin{thm}
  \label{thm:q-main}
  For a distribution $P$ on the input alphabet, one can achieve the commitment rate
  \begin{equation}
    \label{eq:q-achievable}
    H(P)-\chi\bigl(\{P(x);W_x\}\bigr),
  \end{equation}
  with the Holevo mutual information~\cite{holevo}
  $$\chi\bigl(\{P(x);W_x\}\bigr)=S\left( \sum_x P(x)W_x \right)-\sum_x P(x) S(W_x),$$
  where $S(\rho)=-\tr\rho\log\rho$ is the von Neumann entropy of a state.
  \par
  The maximum of the expression~(\ref{eq:q-achievable}) is optimal in the
  case of \emph{no noiseless side communication} during the commit phase.
\end{thm}
\begin{beweis}[(Sketch)]
  For the achievability one proves a coding result similar to proposition~\ref{prop:code},
  with the $\|\cdot\|_1$--norm denoting trace norm. The most crucial point is property $2$:
  our proof used two things: restricting the distributions $W^n_{x^n}$ to typical
  sequences --- this can be done also for states by constructing typical subspaces ---
  and Chernoff bound to obtain a ``small sample''. We use an analogue of this
  for operators from~\cite{Ahlswede:Winter}, stated below as lemma~\ref{lemma:op:chernoff}.
  This technique is actually used in the work of Cai and Yeung~\cite{Cai:Yeung}
  to construct codes for the quantum wiretap channel
  \par
  For the optimality, it is not hard to prove the quantum analogues of eqs.~(A')
  and~(B'), and then the upper bound follows exactly as in our proof of
  proposition~\ref{prop:upper}.
\end{beweis}
\par
\begin{lemma}[Ahlswede,~Winter~\cite{Ahlswede:Winter}]
  \label{lemma:op:chernoff}
  Let $X_1,\ldots,X_L$ be i.i.d.~random variables taking values in the operators
  ${\cal B}({\cal H})$ on the $D$--dimensional Hilbert space ${\cal H}$,
  $0\leq X_\ell\leq \1$, with $A=\E X_\ell\geq\alpha\1$, and let $\eta>0$. Then
  $$\Pr\left\{ \frac{1}{L}\sum_{\ell=1}^L X_\ell \not\in [(1-\eta)A;(1+\eta)A] \right\}
                                          \leq 2D \exp\left( -L\frac{\alpha\eta^2}{2\ln 2} \right),$$
  where $[A;B]=\{X: A\leq X\leq B\}$ is an interval in the operator order.
  \qed
\end{lemma}
\begin{expl}
  \label{expl:bit commitment}
  Assume any set of distinct pure qubit states $W_x=\ketbra{\psi_x}$.
  Then, with $\rho=\sum_x P(x)W_x$, the rate
  $$\max_P \{ H(P)-S(\rho) \}$$
  is achievable. Because of $S(\rho) \leq 1$ this is positive if the input
  alphabet has at least three symbols; in the case of two input symbols it is
  positive iff the two states are non--orthogonal.
  \par
  This is no contradiction to Mayer's no--go theorem for quantum bit commitment
  even though the channel might appear to be noiseless: it is, however, not a noiseless
  qubit channel, because the states are restricted to a set of pure states.
  Modelled as a completely positive map, it would be a measurement--prepare
  channel of the form
  \begin{align*}
    W: {\cal B}\bigl(\C{\cal X}\bigr) &\longrightarrow {\cal B}({\cal H})             \\
                               \sigma &\longmapsto     \sum_x \bra{x}\sigma\ket{x} W_x.
  \end{align*}
  I.e., its use involves a ``guaranteed (von Neumann) measurement'' on all
  messages which come from Alice.
\end{expl}
\par
Regarding theorem~\ref{thm:q-main}, we conjecture the achievable rate stated there
to remain optimal even if unlimited noiseless quantum communication is allowed.
There is however the much more interesting question of more general quantum channels,
for example a depolarising qubit channel, the quantum analogue of a binary symmetric channel:
does it allow bit commitment, and if so, at which rate?
\par
This generalisation may be significant because first of all, information theoretically
secure bit commitment is not possible with noiseless quantum communication~\cite{Mayers}.
Here we have shown that it is possible under the assumption of a noisy channel.
This opens the possibility of perhaps \emph{having} bit commitment under
realistic conditions where one can ensure that all available channels are noisy.

\section{Discussion}
\label{sec:discussion}
We have considered bit--string commitment by using a noisy channel and have
characterised the exact capacity for this task by a single--letter formula.
This implies a lower bound on the coin tossing capacity of that channel
by the same formula, which in fact we conjecture to be an equality.
\par
Satisfactory as this result is, it has to be noted that we are not able in
general to provide an explicit protocol: our proof is based on the random coding
technique and shows only existence. What is more, even if one finds a good
code it will most likely be inefficient: the codebook is just the list of $\xi_{a\mu}$.
In this connection we conjecture that the commitment capacity can be achieved
by random linear codes (compare the situation for channel coding!).
It is in any case an open problem to find efficient good codes, even for the
binary symmetric channel. Note that we only demand efficient \emph{encoding} ---
there is no decoding of errors in our scheme, only an easily performed test.
\par
Our scheme is a block--coding method: Alice has to know the whole of her message,
say a bit string, before she can encode. One might want to use our result as a
building block in other protocols which involve committing to bits at various
stages --- then the natural question arises whether there is an ``online''
version which would allow Alice to encode and send bits as she goes along.
\par
In the same direction of better applicability it would be desirable to extend our
results to a more robust notion of channel: compare the work of~\cite{DKS}
where a cheater is granted partial control over the channel characteristics.
Still, the fixed channel is not beyond application: note that it can be simulated
by pre--distributed data from a trusted party via a ``noisy one--time
pad'' (compare~\cite{beaver} and~\cite{rivest}).
\par
Another open question of interest is to determine the reliability function,
i.e., the optimal asymptotic rate of the error $\epsilon+\delta$
(note that implicit in our proposition~\ref{prop:lower} is a lower bound):
it is especially interesting at $R=0$, because
there the rate tells exactly how secure single--bit commitment can be made.
\par
Finally, we have outlined that a class of quantum channels also allows
bit commitment: they even have a commitment capacity of the same form as
the classical result. This opens up the possibility of unconditionally secure
bit commitment for other noisy quantum channels.
\par
We hope that our work will stimulate the search for optimal rates of other
cryptographic primitives, some of which are possible based on noise,
e.g.~oblivious transfer.

\section*{Acknowledgements}
We thank Ning Cai and Raymond W. Yeung for sharing their ideas on the quantum wiretap
channel, and making available to us their manuscript~\cite{Cai:Yeung}. We also
acknowledge interesting discussions with J. M\"uller--Quade and P. Tuyls at an early stage
of this project.
\par
AW is supported by the U.K.~Engineering and Physical Sciences Research Council.
ACAN and HI are supported by the project ``Research and Development of Quantum
Cryptography'' of the Telecommunications Advancement Organisation as part of
the programme ``Research and Development on Quantum Communication Technology''
of the Ministry of Public Management, Home Affairs, Posts and Telecommunications,
Japan.

\appendix

\section{Typical sequences}
\label{app:typical}
This appendix collects some facts about typical sequences used in the main body of
the text. We follow largely the book of Csisz\'{a}r and K\"orner~\cite{csiszar:koerner}.
\par
The fundamental fact we shall use is the following large deviation version of the law of large
numbers:
\begin{lemma}[Chernoff~\cite{chernoff}]
  \label{lemma:chernoff}
  For i.i.d.~random variables $X_1,\ldots,X_N$, with $0\leq X_n\leq 1$ and
  with expectation $\E X_n=p$:
  \begin{align*}
    \Pr\left\{ \frac{1}{N}\sum_{n=1}^N X_n \geq (1+\eta)p \right\}
                                       &\leq \exp\left(-N\frac{p\eta^2}{2\ln 2}\right), \\
    \Pr\left\{ \frac{1}{N}\sum_{n=1}^N X_n \leq (1-\eta)p \right\}
                                       &\leq \exp\left(-N\frac{p\eta^2}{2\ln 2}\right).
  \end{align*}
  \qed
\end{lemma}
\par
For a probability distribution $P$ on ${\cal X}$ and $\epsilon>0$ define
the set of $\epsilon$--typical sequences:
$${\cal T}^n_{P,\epsilon} = \left\{ x^n : \forall x\ \bigl|N(x|x^n)-P(x)n\bigr|\leq \epsilon n
                                                       \ \&\ P(x)=0\Rightarrow N(x|x^n)=0 \right\},$$
with the number $N(x|x^n)$ denoting the number of letters $x$ in the word $x^n$.
The probability distribution $P_{x^n}(x)=\frac{1}{n}N(x|x^n)$ is called the
\emph{type} of $x^n$. Note that $x^n\in{\cal T}^n_{P,\epsilon}$ is equivalent to
$|P_{x^n}(x)-P(x)| \leq \epsilon$ for all $x$.
\par
These are the properties of typical sequences we shall need:
\begin{equation}
  \label{eq:typ:probability}
  P^{\otimes n}\left( {\cal T}^n_{P,\epsilon} \right)
                                \geq 1-2|{\cal X}|\exp\bigl( -n \epsilon^2/2 \bigr).
\end{equation}
This is an easy consequence of the Chernoff bound, lemma~\ref{lemma:chernoff},
applied to the indicator variables $X_k$ of the letter $x$ in position $k$ in $X^n$,
with $\eta=\epsilon P(x)^{-1}$.
\begin{equation}
  \label{eq:typ:value}
  \forall x^n\in{\cal T}^n_{P,\epsilon}\quad
                                \begin{cases}
                                  P^{\otimes n}(x^n) \leq \exp\bigl( -nH(P)+n\epsilon D \bigr), & \\
                                  P^{\otimes n}(x^n) \geq \exp\bigl( -nH(P)-n\epsilon D \bigr), &
                                \end{cases}
\end{equation}
with the constant $D=\sum_{x:P(x)\neq 0} -\log P(x)$. See~\cite{csiszar:koerner}.
\begin{align}
  \label{eq:typ:upper}
  \left| {\cal T}^n_{P,\epsilon} \right| &\leq \exp\bigl( nH(P)+n\epsilon D \bigr), \\
  \label{eq:typ:lower}
  \left| {\cal T}^n_{P,\epsilon} \right| &\geq \Bigl( 1-2|{\cal X}|\exp\bigl( -n \epsilon^2/2 \bigr) \Bigr)
                                               \exp\bigl( nH(P)-n\epsilon D \bigr).
\end{align}
This follows from eq.~(\ref{eq:typ:value}). These estimates also allow to lower bound
the size of sets with large probability: assume $P^{\otimes n}({\cal C})\geq \eta$, then
\begin{equation}
  \label{eq:shadow:lower}
  |{\cal C}| \geq \Bigl( \eta-2|{\cal X}|\exp\bigl( -n \epsilon^2/2 \bigr) \Bigr)
                                               \exp\bigl( nH(P)-n\epsilon D \bigr).
\end{equation}
\par\medskip
We also use these notions in the ``non--stationary'' case: consider a channel
$W:{\cal X}\longrightarrow{\cal Z}$, and an input string $x^n\in{\cal X}^n$.
Then define, with $\epsilon>0$, the set of conditional $\epsilon$--typical
sequences:
\begin{equation*}\begin{split}
  {\cal T}^n_{W,\epsilon}(x^n) &= \Bigl\{ z^n:\forall x,z\ \bigl|N(xz|x^n z^n)-nW(z|x)P_{x^n}(x)\bigr|
                                                                                 \leq \epsilon n \Bigr. \\
                               &\phantom{===========}
                                \ \&\ \Bigl. W(z|x)=0\Rightarrow N(xz|x^n z^n)=0 \Bigr\}                \\
                               &= \prod_x {\cal T}^{{\cal I}_x}_{W_x,\epsilon{P_{x^n}(x)}^{-1}},
\end{split}\end{equation*}
with the sets ${\cal I}_x$ of positions in the word $x^n$ where $x_k=x$.
The latter product representation allows to easily transport all of the above relations
for typical sequences to conditional typical sequences:
\begin{equation}
  \label{eq:c-typ:probability}
  W^n_{x^n}\left( {\cal T}^n_{W,\epsilon}(x^n) \right)
                                \geq 1-2|{\cal X}||{\cal Z}|\exp\bigl( -n \epsilon^2/2 \bigr).
\end{equation}
\begin{equation}
  \label{eq:c-typ:value}
  \forall x^n\in{\cal T}^n_{W,\epsilon}(x^n)\quad
                               \begin{cases}
                                 W^n_{x^n}(x^n) \leq \exp\bigl( -nH(W|P_{x^n})+n\epsilon E \bigr), & \\
                                 W^n_{x^n}(x^n) \geq \exp\bigl( -nH(W|P_{x^n})-n\epsilon E \bigr), &
                               \end{cases}
\end{equation}
with $E=\max_x \sum_{z:W_x(z)\neq 0} -\log W_x(z)$ and the conditional entropy
$H(W|P)=\sum_x P(x) H(W_x)$.
\begin{align}
  \label{eq:c-typ:upper}
  \left| {\cal T}^n_{W,\epsilon}(x^n) \right|
                                &\leq \exp\bigl( nH(W|P_{x^n})+n\epsilon E \bigr), \\
  \label{eq:c-typ:lower}
  \left| {\cal T}^n_{W,\epsilon}(x^n) \right|
                                &\geq \Bigl( 1-2|{\cal X}||{\cal Z}|\exp\bigl( -n \epsilon^2/2 \bigr) \Bigr)
                                                                  \exp\bigl( nH(W|P_{x^n})-n\epsilon E \bigr).
\end{align}
A last elementary property: for $x^n$ of type $P$ and output distribution $Q$,
with $Q(z)=\sum_x P(x)W_x(z)$,
\begin{equation}
  \label{eq:c-typ:in:typ}
  {\cal T}^n_{W,\epsilon}(x^n) \subset {\cal T}^n_{Q,\epsilon|{\cal X}|}.
\end{equation}
\par\medskip
As an application, let us prove the following lemma:
\begin{lemma}
  \label{lemma:distinct-types}
  For words $x^n$ and $y^n$ with $d_H(x^n,y^n)\geq \sigma n$, and a channel $W$
  such that
  $$\forall x\in{\cal X},P\text{ p.d. with }P(x)=0\quad
                      \left\| W_x-\sum_y P(y)W_y \right\|_1 \geq \eta,$$
  one has, with $\epsilon=\frac{\sigma^2\eta}{2|{\cal X}|^2 |{\cal Z}|}$,
  $$W^n_{y^n}\left( {\cal T}^n_{W,\epsilon}(x^n) \right) \leq 2\exp(-n \epsilon^4/2)$$
\end{lemma}
\begin{beweis}
  There exists an $x$ such that the word $x^{{\cal I}_x}$ (composed of letters $x$ only)
  has distance at least $\frac{1}{|{\cal X}|}\sigma n$ from $y^{{\cal I}_x}$.
  In particular, $N_x:=N(x|x^n)=|{\cal I}_x|\geq \frac{1}{|{\cal X}|}\sigma n$.
  \par
  This implies also, by assumption on the channel,
  \begin{equation*}
    \left\| \frac{1}{N_x}\sum_{k\in{\cal I}_x} W_{y_k} - W_x \right\|_1
                                                    \geq \frac{1}{|{\cal X}|}\sigma\eta.
  \end{equation*}
  Hence there must be a $z\in{\cal Z}$ with
  \begin{equation*}
    \left| \frac{1}{N_x}\sum_{k\in{\cal I}_x} W_{y_k}(z) - W_x(z) \right|
                                         \geq \frac{1}{|{\cal X}| |{\cal Z}|}\sigma\eta.
  \end{equation*}
  By definition, this in turn implies that for all $z^n\in{\cal T}^n_{W,\epsilon}(x^n)$,
  $$\left| N\left(z|z^{{\cal I}_x}\right)-\sum_{k\in{\cal I}_x} W_{y_k}(z) \right|
                                      \geq \frac{1}{2|{\cal X}| |{\cal Z}|}\sigma\eta N_x.$$
  Introducing the sets ${\cal J}_{xy}=\{ k\in{\cal I}_x : y_k=y \}$, with cardinalities
  $N_{xy}=|{\cal I}_{yx}|$, there is a $y$ such that (still for all $z^n\in{\cal T}^n_{W,\epsilon}(x^n)$),
  \begin{equation*}\begin{split}
    \left| N\left(z|z^{{\cal J}_{xy}}\right)-N_{xy}W_y(z) \right|
                                      &\geq \frac{1}{2|{\cal X}|^2 |{\cal Z}|}\sigma\eta N_x   \\
                                      &\geq \frac{1}{2|{\cal X}|^2 |{\cal Z}|}\sigma\eta N_{xy}.
  \end{split}\end{equation*}
  This implies
  $$N_{xy} \geq \frac{1}{4|{\cal X}|^2 |{\cal Z}|}\sigma\eta N_x
          \geq \frac{1}{4|{\cal X}|^3 |{\cal Z}|}\sigma^2\eta n,$$
  and with lemma~\ref{lemma:chernoff} we obtain the claim.
\end{beweis}

\end{document}